\documentclass{Interspeech2024}
\interspeechcameraready 
\usepackage{multicol,multirow}
\usepackage{url,cite}
\usepackage{xcolor}
\usepackage{tabularx,lipsum,amssymb,amsmath,pifont,graphicx,adjustbox}
\usepackage{listings}
\usepackage{hyperref}
\usepackage{lipsum}
\usepackage[labelfont=bf]{caption} 
\hypersetup{
    colorlinks=true,
    linkcolor=purple,
    filecolor=magenta,      
    urlcolor=purple,
}
\newcommand{\newpara}[1]{\vspace{2pt}\noindent\textbf{#1}}
\newcommand{\cmark}{\ding{51}}
\newcommand{\xmark}{\ding{55}}
\newcolumntype{Y}{>{\centering\arraybackslash}X}
\definecolor{codegreen}{rgb}{0,0.6,0}
\definecolor{codegray}{rgb}{0.5,0.5,0.5}
\definecolor{codepurple}{rgb}{0.58,0,0.82}
\definecolor{backcolour}{rgb}{0.95,0.95,0.92}

\lstdefinestyle{mystyle}{
    backgroundcolor=\color{backcolour},   
    commentstyle=\color{codegreen},
    keywordstyle=\color{magenta},
    numberstyle=\tiny\color{codegray},
    stringstyle=\color{codepurple},
    basicstyle=\ttfamily\scriptsize, 
    breakatwhitespace=false,        
    xleftmargin=0.4cm,
    breaklines=true,                 
    captionpos=b,                    
    keepspaces=true,                 
    numbers=left,                    
    numbersep=5pt,                  
    showspaces=false,                
    showstringspaces=false,
    showtabs=false,                  
    tabsize=2
}

\title{ESPnet-SPK: full pipeline speaker embedding toolkit with reproducible recipes, self-supervised front-ends, and off-the-shelf models}
\name[affiliation={1,*}]{Jee-weon}{Jung}
\name[affiliation={1,*}]{Wangyou}{Zhang}
\name[affiliation={1,*}]{Jiatong}{Shi}
\name[affiliation={2}]{Zakaria}{Aldeneh}
\name[affiliation={2}]{Takuya}{Higuchi}
\name[affiliation={1}]{\\Alex}{Gichamba}
\name[affiliation={2}]{Barry-John}{Theobald}
\name[affiliation={2}]{Ahmed Hussen}{Abdelaziz}
\name[affiliation={1}]{Shinji}{Watanabe}

\address{
  $^1$Carnegie Mellon University, USA\qquad
  $^2$Apple, USA}
\email{jeeweonj@ieee.org, shinjiw@ieee.org}


\begin{document}
\maketitle
\begin{abstract}
This paper introduces ESPnet-SPK, a toolkit designed for training and utilizing speaker embedding extractors. It offers an open-source platform, facilitating effortless construction of models ranging from the x-vector to the SKA-TDNN, thanks to its modular architecture that simplifies the development of variants. The toolkit advances the use of speaker embeddings across various tasks where outdated embeddings are often employed, enabling the broader research community to use advanced speaker embeddings effortlessly. Pre-trained extractors are readily available for off-the-shelf use. The toolkit also supports integration with various self-supervised learning features. ESPnet-SPK features over 30 recipes: seven speaker verification recipes, including reproducible WavLM-ECAPA with an EER of 0.39\% on the Vox1-O benchmark and diverse downstream tasks, including text-to-speech and target speaker extraction. It even supports speaker similarity evaluation for singing voice synthesis and more.

\end{abstract}

\section{Introduction}
\label{sec:intro}

In recent years, the rapid advancements in speech technologies, including speaker recognition with deep learning, have yielded robust speaker embeddings~\cite{snyder2018x,jung2019rawnet,wan2018generalized}.
These compact representations of unique speaker characteristics are pivotal in diverse tasks, including speaker verification, diarization~\cite{park2022review,jung2023search}, Text-To-Speech (TTS)~\cite{YourTTS-Casanova2022}, and other speech-related tasks~\cite{VoiceFilter-Wang2019,SpEx-Xu2020,huang2021far,huang2023singing}. 
With growing demand for speaker embedding extractors, numerous open-source toolkits have emerged to streamline the training and deployment of these models~\cite{Kaldi-Povey2011,chung2020defence,wang2023wespeaker,tong2021asv,zhang2022paddlespeech,ravanelli2021speechbrain,kuchaiev2019nemo,zheng20233d}.

Table~\ref{tab:toolkits} shows prominent speaker recognition toolkits.
Though these toolkits have substantially contributed to the community, they often fail to address its rapidly evolving requirements.
Many lack a comprehensive set-up that integrates data preprocessing, feature extraction, and training/inference design, limiting researchers' access to state-of-the-art (SOTA) models. Consequently, inferior speaker embeddings are frequently used in downstream tasks~\cite{YourTTS-Casanova2022, hayashi2021espnet2}. 
Furthermore, despite the proven effectiveness of the features of Self-Supervised Learning (SSL) models~\cite{yang21c_interspeech, chen2022wavlm}, recipes for incorporating these features into speaker embedding extractor development are notably absent.

To complement these existing toolkits, we introduce ESPnet-SPK, a new open-source toolkit for speaker embedding extraction designed to meet the research community's evolving needs. 
ESPnet-SPK aims to address multiple objectives: foremost, it offers a unified platform for speaker recognition researchers to develop, compare, and evaluate models across diverse datasets in a unified environment.
It features a wide array of models, from the x-vector~\cite{snyder2018x} to the advanced SKA-TDNN~\cite{mun2023frequency}. Its modular composition facilitates novel model development; the combination of modules can result in hundreds of different models. Additionally, it supports evaluations for the newly emerging Spoofing-robust Automatic Speaker Verification (SASV) task~\cite{jung2022sasv}, highlighting the toolkit's scalability and adaptability.
\begin{table}[t!]
  \caption{Open-source speaker verification toolkits. $\triangle$: pre-trained models exist, but not prepared for off-the-shelf usage. $^\dagger$: currently does not support a SSL feature-based speaker verification recipe, but possible with modifications.}
  \label{tab:toolkits}
  \centering
  \adjustbox{max width=\linewidth}{
  \begin{tabular}{lccc}
    \toprule
    \textbf{Name} & \textbf{Datasets} & \textbf{SSL front-end} &  \textbf{Off-the-shelf}\\
    \midrule

    \href{https://github.com/Snowdar/asv-subtools}{ASV-subtools}\cite{tong2021asv} & Multiple & \xmark & \xmark\\
    \href{https://github.com/NVIDIA/NeMo}{NeMo}\cite{kuchaiev2019nemo} & VoxCeleb & \xmark & \xmark\\
    \href{https://github.com/kaldi-asr/kaldi}{Kaldi}\cite{Kaldi-Povey2011} & Multiple & \xmark & $\triangle$\\
    \href{https://github.com/clovaai/voxceleb_trainer}{VoxCeleb\_Trainer}\cite{chung2020defence} & VoxCeleb & \xmark & $\triangle$\\
    \href{https://github.com/wenet-e2e/wespeaker}{Wespeaker}\cite{wang2023wespeaker} & Multiple & \xmark & \cmark\\
    \href{https://github.com/alibaba-damo-academy/3D-Speaker/tree/main/egs}{3D-Speaker}\cite{zheng20233d} & Multiple & \xmark & $\triangle$ \\
    \href{https://github.com/PaddlePaddle/PaddleSpeech}{PaddleSpeech}\cite{zhang2022paddlespeech} &  VoxCeleb & \cmark$^\dagger$ & \cmark \\
    \href{https://github.com/speechbrain/speechbrain}{SpeechBrain}\cite{ravanelli2021speechbrain} & VoxCeleb & \cmark$^\dagger$ & \cmark\\
    \toprule
    \href{https://github.com/espnet/espnet}{ESPnet-SPK} & Multiple & \cmark & \cmark\\
    \bottomrule
  \end{tabular}
  }
\end{table}

Moreover, the toolkit grants the broader speech community easy access to the latest models for their research, providing pre-trained speaker embedding extractors and the necessary tools for effortless ``off-the-shelf'' applications. 
Speaker embeddings generated by ESPnet-SPK are versatile and {\em can be applied to a wide range of speech-processing tasks beyond speaker verification}; it currently supports over $30$ recipes on diverse downstream tasks, including TTS, TSE, target speech recognition, and more. 
We demonstrate its utility in two specific tasks with reproducible ESPnet recipes: multi-speaker TTS and Target Speaker Extraction (TSE).

Additionally, ESPnet-SPK incorporates SSL features into speaker embedding development, enhancing task performance. 
This integration augments the {\em number of available model architectures further to thousands}, utilizing S3PRL\footnote{\url{https://github.com/s3prl/s3prl}.}. 
We showcase a recipe with an Equal Error Rate (EER) of 0.39\% on the Vox1-O evaluation protocol, utilizing WavLM-Large~\cite{chen2022wavlm} and ECAPA-TDNN~\cite{desplanques2020ecapa}.
This capability distinguishes ESPnet-SPK from other toolkits, like PaddleSpeech and SpeechBrain~\cite{ravanelli2021speechbrain}, offering broader and more competitive model support beyond a single recipe and exemplifying its potential to meet and exceed the evolving benchmarks of the research community.

\section{Framework}
\label{sec:framework}
\subsection{Design}
\label{ssec:design}
\begin{figure}[t]
    \centering
    \includegraphics[width=\linewidth]{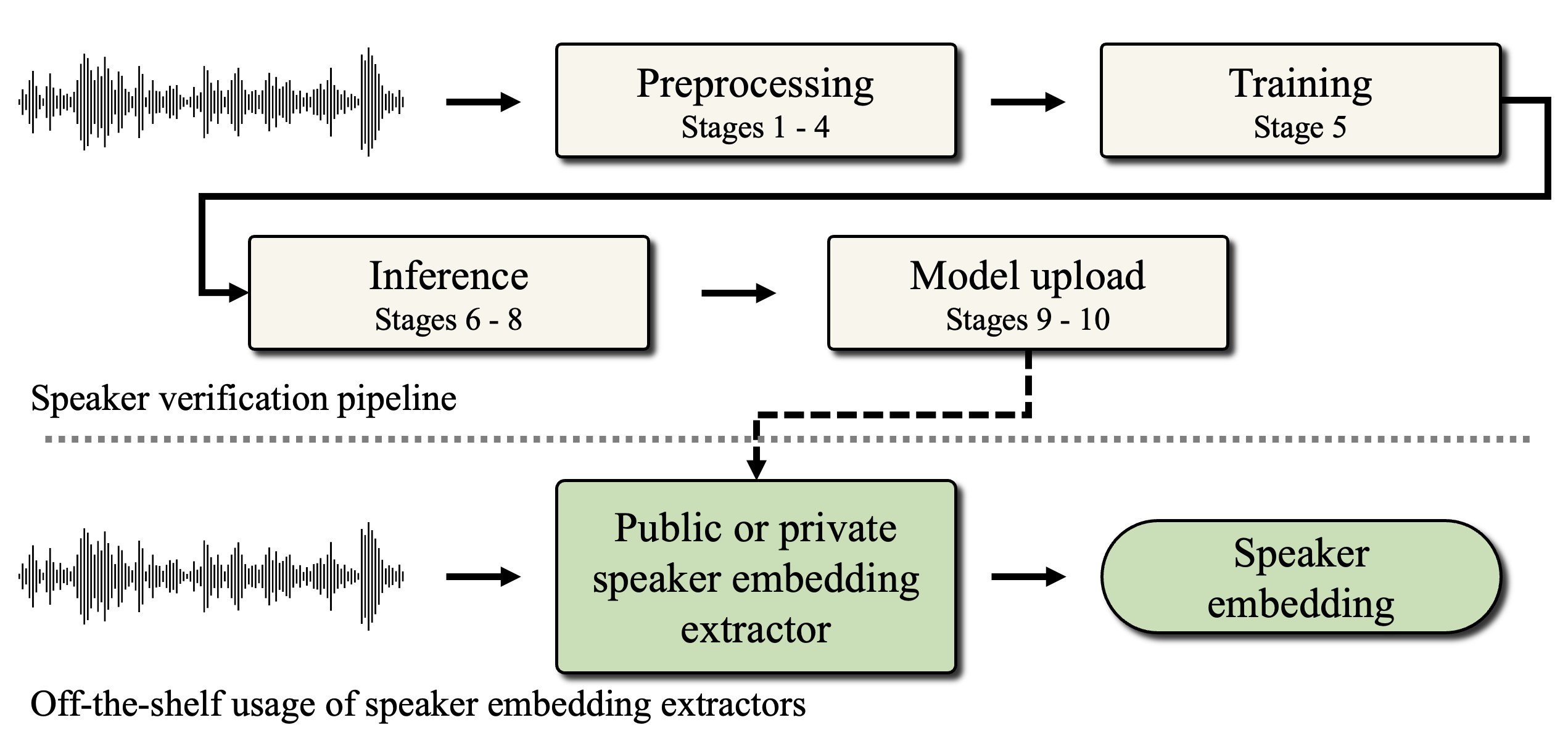}
    \caption{
    Process Pipeline of ESPnet-SPK, structured in multiple stages akin to the Kaldi~\cite{Kaldi-Povey2011} speech processing toolkit. The top section outlines stages 1 through 10, the speaker verification process, with stages 9 and 10 dedicated to optionally publishing trained speaker embedding extractors. Furthermore, it highlights the ease of using publicly available embedding extractors in an off-the-shelf manner.}
    \label{fig:overall}
\end{figure}

The ESPnet-SPK toolkit consists of two main components: the speaker verification recipe and off-the-shelf speaker embedding extraction. 
The speaker verification recipe comprises eight stages targeted to help speaker recognition researchers and detailed in Section \ref{ssec:sv_recipe}. 
The off-the-shelf component, outlined in Section~\ref{ssec:off-the-shelf}, addresses the deployment of trained speaker embedding extractors, making them readily available for public use.
This sub-section is specifically designed to support researchers outside the speaker recognition field on top of those within it, illustrating the framework's comprehensive approach. Figure~\ref{fig:overall} provides an overview of the entire toolkit.

\subsection{Speaker verification recipe}
\label{ssec:sv_recipe}
\newpara{Preprocessing.}
The preprocessing phase is divided into four stages, each with a distinct function:
\begin{itemize}
    \item Stage 1: Downloading and preparing datasets, including those for augmentation;
    \item Stage 2: Applying speed perturbation to audio files~\cite{Audio-Ko2015};
    \item Stage 3: Formatting the data for processing;
    \item Stage 4: Calculating relevant statistics.
\end{itemize}
These initial stages are consistent with the broader ESPnet framework, and we direct the readers to Watanabe \emph{et al.}~\cite{watanabe2018espnet} for comprehensive details.
A notable distinction in our approach, as inspired by Yamamoto \emph{et al.}~\cite{yamamoto2019speaker} and Chen \emph{et al.}~\cite{chen2023build}, is treating speakers in speed-perturbed utterances as distinct entities rather than identical ones. This augmentation method has proven beneficial for enhancing speaker verification performance.

\newpara{Training.}
Stage 5 executes the core training process.
Through a centralized \texttt{yaml}-style configuration file, we specify the model architecture, objective function, optimizer, and additional training hyperparameters. The configuration file is then used to train the speaker embedding extractor efficiently.

\newpara{Supported models.}
ESPnet-SPK currently supports five 
predefined model architectures: x-vector~\cite{snyder2018x}, MFA-Conformer~\cite{zhang2022mfa}, ECAPA-TDNN~\cite{desplanques2020ecapa}, RawNet3~\cite{jung2022pushing}, and SKA-TDNN~\cite{mun2023frequency}. The number of architectures, nevertheless, can increase to thousands with different module combinations, SSL front-ends, and encoders from the other ESPnet ecosystem.
While we omit detailed architectural descriptions due to space constraints, we direct readers to the respective publications for in-depth information.

\begin{figure}[t!]
    \centering
    \includegraphics[width=\linewidth]{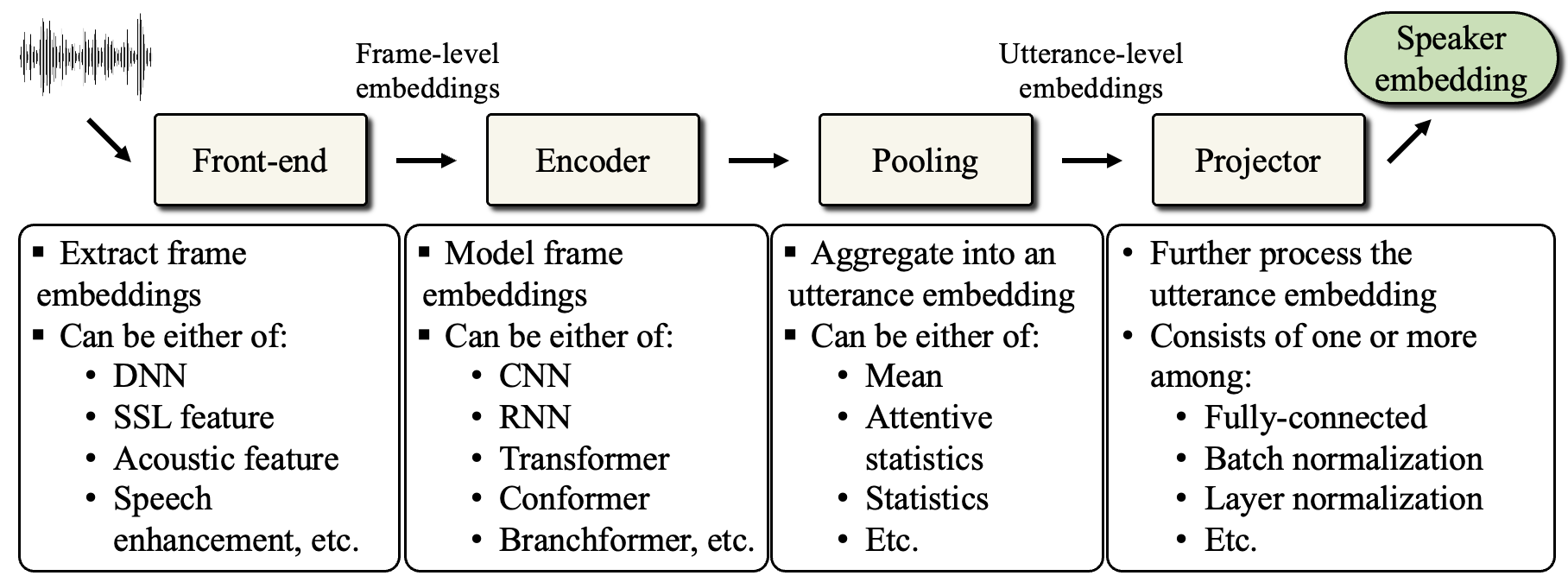}
    \caption{Illustration of the modular sub-components of the speaker embedding extractor. Users can effortlessly construct thousands of model architectures in the configuration file by combining these sub-components.}
    \label{fig:components}
\end{figure}

\newpara{Modularized sub-components.}
ESPnet-SPK organizes the speaker embedding extractor into four key sub-components as illustrated in Figure~\ref{fig:components}: front-end, encoder, pooling, and projector.
The {\em front-end} processes raw waveforms to generate a sequence of input features.
For models like RawNet3 that directly process raw waveforms, the front-end includes layers such as Sinc-conv with normalization to provide an integrated format for the encoder input.
The {\em encoder} is the principal block for frame-level feature modeling, e.g., convolutional or Conformer blocks.
The {\em pooling} contains aggregation methods, including the widely used attentive statistics pooling~\cite{okabe2018attentive}.
The {\em projector} consists of additional layers, typically including fully-connected, batch normalization~\cite{ioffe2015batch}, or layer normalization layers~\cite{ba2016layer}.

\newpara{Objective functions, optimizers, and learning rate schedulers.}
ESPnet-SPK accommodates a wide array of objective functions, from traditional cross-entropy to advanced AAM-Softmax~\cite{deng2019arcface}, enhanced with sub-center and inter top-k techniques~\cite{zhao2021speakin}.
It shares the same optimizers and learning rate schedulers as the broader ESPnet ecosystem, ensuring access to various established configurations.

\newpara{Inference.}
\label{ssec:inference}
The inference phase is segmented into stages 6, 7, and 8, each serving a specific purpose:
\begin{itemize}
    \item Stage 6: Extracting speaker embeddings from all utterances within the test set;
    \item Stage 7: Applying post-normalization and enhancement processes at the score level;
    \item Stage 8: Calculating the metrics.
\end{itemize}

\textbf{Stage 6} additionally extracts embeddings for utterances used in subsequent processing in Stage 7.
This stage is crucial for preparing cohort sets for score normalization and quality measurement. 
\textbf{Stage 7} applies additional processing in the score level. Once all trial scores are derived, configurations determine whether to apply score normalization and quality measure functions.
For score normalization, we support AS-norm~\cite{matejka2017analysis}; for quality measure function, we implement those mentioned in \cite{torgashov2023id}.
\textbf{Stage 8} lastly calculates the metrics, EER, and minimum Detection Cost Function (minDCF). 
For the minDCF, we use a $p_{\text{trg}}$, ${C_{\text{false\_alarm}}}$, and $C_{\text{miss}}$ of $0.05$, $1$, and $1$ respectively.

\begin{figure}
    \centering
\begin{lstlisting}[language=Python]
import numpy as np
from espnet2.bin.spk_inference import Speech2Embedding

# from uploaded models
speech2spk_embed = Speech2Embedding.from_pretrained(model_tag="espnet/voxcelebs12_rawnet3")
embedding = speech2spk_embed(np.zeros(16500))

# from checkpoints trained by oneself
speech2spk_embed = Speech2Embedding(model_file="model.pth", train_config="config.yaml")
embedding = speech2spk_embed(np.zeros(32000))
\end{lstlisting}
    \caption{Sample Code demonstrating the use of public and custom speaker embedding extractors of ESPnet-SPK.}
    \label{fig:shelf}
\end{figure}

\subsection{Off-the-shelf usage}
\label{ssec:off-the-shelf}
After completing the training and inference phases, models can be shared publicly through stages 9 and 10:

\begin{itemize}
    \item Stage 9: Prepares the model for distribution;
    \item Stage 10: Uploads the model to Huggingface.
\end{itemize}

For the models already made publicly available, anyone can use them off the shelf.
Users can easily upload their own models by simply specifying the model name.\footnote{Available list of models live at \url{https://github.com/espnet/espnet/egs2/[corpus]/spk1/README.md}.} 
This functionality allows individuals within and outside the speaker recognition community to utilize or contribute to the ESPnet-SPK ecosystem easily. 
Figure~\ref{fig:shelf} demonstrates extracting a speaker embedding from a pre-trained RawNet3 model in under ten lines of Python code, showcasing the framework's user-friendliness.

\section{Experiments}
\label{sec:exp}
ESPnet-SPK has been developed using PyTorch and lives at \texttt{\url{https://github.com/espnet/espnet}}.

\subsection{Corpus}
\label{ssec:corpus}
This paper investigates the widely used VoxCeleb 1\&2~\cite{nagrani2017voxceleb,chung2018voxceleb2}, the newly introduced VoxBlink~\cite{lin2024voxblink}, and the ASVspoof 2019~\cite{wang2020asvspoof} corpora. Although the toolkit features several additional recipes, including VOiCES and the SdSV Challenge, this paper does not cover them due to space constraints. 

\newpara{VoxCeleb 1\&2} includes $7,333$ speakers with over $1.2$M utterances. Our experiments primarily utilize the combined development sets, comprising approximately $2,690$ hours of speech from $7,205$ speakers. 
The Vox1-O evaluation protocol, involving around $37$k trials from $40$ speakers in the VoxCeleb1 test set, serves as our primary benchmark.

\newpara{VoxBlink} comprises two subsets: the full set, with 1.45 million utterances from 38k speakers, and the clean set, a subset with 1.02 million utterances from 18k speakers.
VoxBlink complements VoxCeleb by focusing on a broader demographic from YouTube videos, offering a contrast to VoxCeleb’s celebrity focus.

\newpara{ASVspoof 2019} logical access test set is used to show the Spoofing-robust Automatic Speaker Verification (SASV)~\cite{jung2022sasv} capabilities of the embedding extractors, highlighting the toolkit's adaptability to recent trends in speaker verification.

\subsection{Vox1-O and ASVspoof 2019 benchmarks}
\label{ssec:mainres}
Table~\ref{tab:mainres} displays the benchmark results across various representative models using the Vox1-O and SASV 2022 evaluation protocols. Input features were primarily mel-spectrograms derived from 3 seconds of speech, except for RawNet3, which processes raw waveforms directly. We adopted the AAM-softmax with sub-center, and inter top-k penalty loss functions~\cite{zhao2021speakin} with an Adam optimizer.

In the Vox1-O benchmark, SKA-TDNN and RawNet3 emerged as top performers among the five models presented, all accessible off-the-shelf. The SASV 2022 benchmark revealed the vulnerability of contemporary speaker verification systems to spoofing attacks, underscoring the critical need for future research in SASV. RawNet3 demonstrated the highest resistance to spoofing among the evaluated models, which is consistent with studies in audio anti-spoofing, where raw waveform-based models demonstrate superior performance~\cite{jung2022aasist}.

\begin{table}[t!]
  \caption{Comparison of models on the Vox1-O and SASV 2022 evaluation protocols. Lower is better for all three metrics.}
  \label{tab:mainres}
  \centering
  \resizebox{\columnwidth}{!}{%
  \begin{tabular}{lccc}
    \toprule
    \textbf{Model} & \textbf{EER~(\%) $\downarrow$} & \textbf{minDCF $\downarrow$} & \textbf{SASV-EER~(\%) $\downarrow$}\\
    \midrule
    x-vector~\cite{snyder2018x} & 1.81 & 0.1251 & 25.84\\
    MFA-Conformer~\cite{zhang2022mfa} & 0.86 & 0.0627 & 24.71\\
    ECAPA-TDNN~\cite{desplanques2020ecapa} & 0.85 & 0.0666 & 26.12\\
    RawNet3~\cite{jung2022pushing} & 0.73 & 0.0581 & \textbf{17.41}\\
    SKA-TDNN~\cite{mun2023frequency} & \textbf{0.72} & \textbf{0.0457} & 21.75\\
    \bottomrule
  \end{tabular}
  }
\end{table}
\begin{table}[t!]
  \caption{Comparison of ECAPA- and SKA-TDNN performance using different front-ends on the Vox1-O evaluation protocol. The table compares mel-spectrogram, WavLM (fixed), and WavLM (jointly fine-tuned). Reported in EERs (\%). ``Mel-spec'': mel-spectrogram, ``tuned'': jointly fine-tuned.}
  \label{tab:sslres}
  \centering
  \resizebox{\columnwidth}{!}{%
  \begin{tabular}{lccc}
    \toprule
    \multirow{2}{*}{\textbf{Model}} & \multicolumn{3}{c}{\textbf{Front-end}} \\
    & \textbf{Mel-spec} & \textbf{WavLM (fixed)} & \textbf{WavLM (tuned)}\\
    \midrule
    ECAPA-TDNN & 0.85 & 0.60 & \textbf{0.39}\\
    SKA-TDNN & 0.72 & 0.56 & 0.51\\
    \bottomrule
  \end{tabular}%
  }

\end{table}

\subsection{SSL front-end replacements}
\label{ssec:sslres}
Table~\ref{tab:sslres} presents the impact of substituting the front-end from mel-spectrogram to WavLM-Large SSL model. The integration with S3PRL enables straightforward access to over 100 SSL models through minor adjustments in the \texttt{yaml} configuration file. WavLM was chosen for its exceptional performance in speaker-related tasks, as demonstrated in the SUPERB benchmark~\cite{yang21c_interspeech}. Our experiments used ECAPA-TDNN and SKA-TDNN in three configurations: traditional mel-spectrogram front-end, fixed WavLM front-end, and jointly fine-tuned WavLM front-end, keeping all other hyperparameters consistent with Section~\ref{ssec:mainres}.

The results indicate a consistent trend: WavLM surpasses the mel-spectrogram, and joint fine-tuning of the SSL front-end yields better results than the fixed approach. 
Specifically, for ECAPA-TDNN, ESPnet-SPK achieved an EER of 0.39\%, closely mirroring the 0.38\% EER reported in the original work~\cite{chen2022wavlm}, marking the first reproducible benchmark of this performance. However, for SKA-TDNN, the benefits of jointly fine-tuning WavLM were more modest. This could be due to SKA-TDNN initially processing inputs through the ``fcwSKA Block,'' which could align more effectively with traditional acoustic features with linearly positioned frequency bands.

\subsection{Recipes on VoxBlink}
\label{ssec:vbres}
Table~\ref{tab:vbres} presents the experimental results on the VoxBlink corpus~\cite{lin2024voxblink}.
While the authors released the full- and the clean- sets, this paper marks the first instance of model performance being reported on the full set, highlighting ESPnet-SPK's role as a pioneer in providing open-source recipes.
Our investigation included training with separate sets (rows 2 and 3) and their integration with the VoxCeleb1\&2 development sets (rows 4 and 5).
The data quality proved pivotal; the model trained on the clean set outperformed the full set despite the latter's larger volume of data (2.1k vs 1.6k hours).
Contrary to the findings reported in the original VoxBlink study, where combining VoxCeleb to the training data yielded a 10\% improvement, our results showed superior performance when we trained the model with only VoxCeleb data. This discrepancy could stem from our inclusion of the VoxCeleb1 development set.

\begin{table}[t!]
  \caption{
  Results from experiments using the VoxBlink dataset~\cite{lin2024voxblink}.
  This is the inaugural report of model training exclusively with the VoxBlink-full set. 
  The model architecture is fixed to RawNet3. ``Vb-full'': VoxBlink-full, ``Vb-cln'': VoxBlink-clean.}
  \label{tab:vbres}
  \centering
  \begin{tabularx}{\linewidth}{YYYcc}
    \toprule
    \textbf{Voxcelebs} & \textbf{Vb-full} & \textbf{Vb-cln} & \textbf{EER (\%) $\downarrow$} & \textbf{minDCF $\downarrow$}\\
    \midrule
    \checkmark & - & - & \textbf{0.73} & 0.0581\\
    - & \checkmark & - & 2.68 & 0.1893\\
    - & - & \checkmark & 2.51 & 0.1858\\
    \toprule
    \checkmark & \checkmark & - & 0.78 & 0.0655\\
    \checkmark & - & \checkmark & 0.77 & \textbf{0.0556}\\
    \bottomrule
  \end{tabularx}
  
\end{table}
\begin{table}[t!]
  \caption{
  Results from TTS experiments using the VITS model~\cite{kim2021conditional} on the VCTK corpus~\cite{veaux201x_cstr_vctk}, employing a range of speaker embeddings from ESPnet-SPK. Reference scores for Mean Opinion Score (MOS) and Speaker similarity MOS (SMOS) are $4.04$ and $4.84$, respectively.
  }
  \label{tab:ttsres}
  \centering
  \resizebox{\linewidth}{!}{%
  \begin{tabular}{lcccc}
    \toprule
    \textbf{Embedding} & \textbf{MCD $\downarrow$}  & \textbf{WER (\%) $\downarrow$} & \textbf{MOS $\uparrow$} & \textbf{SMOS $\uparrow$} \\
    \midrule
    x-vector \cite{snyder2018x,ravanelli2021speechbrain} & 6.76 & 9.83 & 3.39 & \textbf{3.92}\\
    Rawnet3~\cite{jung2022pushing}& 6.76 & \textbf{7.67} & \textbf{3.49} & 3.84\\
    WavLM(fixed)+ECAPA & \textbf{6.69} & 9.30 & 3.42 & 3.82\\
    WavLM(tuned)+ECAPA  & 6.75 & 8.04 & 3.45 & 3.85\\
    \bottomrule
  \end{tabular}%
  }
  
  \vspace{-7pt}
\end{table}




\vspace{-5pt}
\subsection{Downstream tasks demonstration}
\label{ssec:ttsres}

\newpara{TTS results.} Adopting different speaker embeddings affects the performance of multi-speaker TTS systems~\cite{hayashi2021espnet2,YourTTS-Casanova2022}. 
Nevertheless, the specific impact of different pre-trained speaker embedding extractors on TTS still needs to be explored.
We evaluate the performance of four pre-trained speaker embeddings by integrating them with a VITS model~\cite{kim2021conditional} on the VCTK corpus~\cite{veaux201x_cstr_vctk}. Notably, the ESPnet ecosystem provides more than $20$ TTS recipes, which can leverage any speaker embedding extractor of ESPnet-SPK, including LibriTTS and AISHELL-3.

Our evaluation uses both objective and subjective metrics, including Mel Cepstral Distortion (MCD), Word Error Rate (WER) assessed by Whisper-small~\cite{radford2023robust}, UTMOS~\cite{saeki22c_interspeech}, Mean Opinion Score (MOS), and Speaker Similarity MOS (SMOS). Each utterance received six votes from English speakers for both MOS and SMOS evaluations.
Results, presented in Table~\ref{tab:ttsres}, illustrate the ease of comparing different speaker embeddings through ESPnet-SPK. RawNet3 performed best in MOS and WER evaluations, while the top systems for other metrics varied.

\newpara{TSE results.}
\label{ssec:tseres}
Speaker embeddings play a crucial role in TSE, where they act as a reference to isolate the target speaker's voice from a mix of multiple speakers. The selection of an optimal speaker embedding is essential for effective TSE. 
We evaluate the Perceptual Evaluation of Speech Quality (PESQ) and Signal-to-Distortion Ratio (SDR) performance of four distinct speaker embeddings in TSE using the WSJ0-2mix benchmark~\cite{Deep-Hershey2016}.\footnote{We also provide a recipe on LibriMix, another popular TSE corpus.} 
We use the popular \texttt{min} version of WSJ0-2mix, where we ensure that each sample contains only fully overlapping speech samples with careful trimming.
We exploit the ESPnet-SE recipe, demonstrating the ease of connectivity within the ESPnet ecosystem. 
Our experiments with the TD-SpeakerBeam model~\cite{Improving-Delcroix2020} involve fusing the speaker embedding after the initial temporal convolutional network block.
Our primary aim is not to achieve peak performance but to demonstrate the straightforward integration of developed speaker embeddings into various related tasks.

The findings, detailed in Table~\ref{tab:tseres}, demonstrate the facility of ESPnet-SPK for comparing various speaker embeddings through ESPnet-SPK in TSE studies.
Despite its relatively lower performance in speaker verification, we discovered that the x-vector (512-dimensional) shows the highest TSE performance across both validation and test sets.
This observation aligns with insights from~\cite{zhao2022probing,liu2023quantitative}.
In contrast, speaker embeddings with 192 dimensions yield inferior results on the test dataset, suggesting vulnerability to target speaker confusion issues.
We analyze that the x-vector's higher dimensionality may encapsulate a broader range of information, potentially harmful to speaker verification yet beneficial for downstream tasks.

\vspace{-5pt}
\subsection{Additional results on loss functions, training data, and architecture}
\label{ssec:ablations}

Lastly, Table~\ref{tab:ablation} presents a series of ablations assessing the impact of varying loss functions, training datasets, and architectures. First, we examine two modifications to RawNet3: altering the objective function and adjusting the training dataset. Substituting the complex sub-center AAM-softmax and inter top-k with only AAM-softmax increased the EER to 0.79\%. Exclusively using VoxCeleb2-dev for training by omitting VoxCeleb1-dev further raised the EER to 0.85\%. These findings underscore the significance of objective functions and training data for optimal performance. Again, the experiments demonstrate the ease of using different loss functions, training data, and architectures within the toolkit. 

Subsequent ablations focused on models featuring SSL front-end features inspired by \cite{peng2023parameter}.
Training exclusively with VoxCeleb1-dev yielded an EER of 1.33\%. Removing the ECAPA-TDNN encoder to streamline the architecture to WavLM, attentive statistics pooling, and a projection layer led to an EER of 1.88\%. These findings affirm the potential of SSL front-ends paired with simpler architectures and datasets for competitive outcomes, and also illustrate ESPnet-SPK's capability to facilitate exploration across various SSL model implementations.

\begin{table}[t!]
  \caption{TSE results on WSJ0-2mix (\texttt{min} mode) using the TD-SpeakerBeam model~\cite{Improving-Delcroix2020} with various speaker embeddings. 
  }
  \label{tab:tseres}
  \centering
  \resizebox{\columnwidth}{!}{%
  \begin{tabular}{lcccc}
    \toprule
    \multirow{2}{*}{\textbf{Embedding extractor}} & \multicolumn{2}{c}{\textbf{PESQ $\uparrow$}} & \multicolumn{2}{c}{\textbf{SDR (dB) $\uparrow$}} \\
    & \textbf{validation} & \textbf{test} & \textbf{validation} & \textbf{test} \\
    \midrule
    x-vector~\cite{snyder2018x,ravanelli2021speechbrain} & \textbf{3.57} & \textbf{3.25} & \textbf{18.96} & \textbf{15.70} \\
    RawNet3~\cite{jung2022pushing} & 3.54 & 2.64 & 18.65 & 9.21 \\
    WavLM-Large(fixed)+ECAPA & 3.53 & 2.77 & 18.60 & 10.60 \\
    WavLM-Large(tuned)+ECAPA & 3.53 & 2.79 & 18.68 & 11.27 \\
    \bottomrule
  \end{tabular}%
  }
\end{table}

\begin{table}[t!]
  \caption{
  Ablations on selected components. 
  Objective function and train data is ablated on RawNet3. Model architecture is ablated on ECAPA-TDNN.}
  \label{tab:ablation}
  \centering
  \adjustbox{max width=\linewidth}{
  \begin{tabular}{lcc}
    \toprule
    \textbf{Model} & \textbf{EER~(\%) $\downarrow$} & \textbf{minDCF $\downarrow$}\\
    \midrule
    RawNet3 & 0.73 & 0.0581\\
    \quad$\rightarrow$ w/o sub-center \& inter top-k & 0.79 & 0.0626\\
    \quad$\rightarrow$ w/o VoxCeleb1 dev & 0.85 & 0.0698\\
    \hline\hline
    WavLM-Large (fixed) + ECAPA & 0.60 & 0.0446\\
    \quad$\rightarrow$ w/o VoxCeleb2 dev & 1.33 & 0.1157\\
    \quad\quad$\rightarrow$ w/ identity encoder & 1.88 & 0.1363\\
    \bottomrule
  \end{tabular}
  }
  
\end{table}

\vspace{-5pt}
\section{Conclusion}
\label{sec:conclusion}
We introduced ESPnet-SPK, a versatile open-source toolkit designed for speaker verification and embedding extraction.
This toolkit provides a fully integrated pipeline featuring various reproducible recipes. It currently supports five of the most popular speaker embedding extractors and facilitates the creation of countless new model variants via its modular architecture and integration with SSL technologies. 
Furthermore, ESPnet-SPK offers a seamless off-the-shelf user experience, enabling a wide range of interdisciplinary applications to easily access and deploy SOTA speaker embeddings for various downstream tasks.

\clearpage
\section{Acknowledgment}
We use PSC Bridges2 and NCSA Delta via ACCESS CIS210014, by National Science Foundation grants \#2138259, \#2138286, \#2138307, \#2137603, and \#2138296.
This work was supported by a grant from Apple, Inc. Any views, opinions, findings, and conclusions or recommendations expressed in this material are those of the authors and should not be interpreted as reflecting the views, policies or position, either expressed or implied, of Apple Inc.
\bibliographystyle{IEEEtran}
\bibliography{shortstrings,mybib}

\begin{thebibliography}{10}

\bibitem{snyder2018x}
D. Snyder, D. Garcia-Romero et~al.,
\newblock ``X-vectors: Robust {DNN} embeddings for speaker recognition,''
\newblock in {\em Proc. ICASSP}, 2018.

\bibitem{jung2019rawnet}
J.-w. Jung et~al.,
\newblock ``{RawNet}: Advanced end-to-end deep neural network using raw waveforms for text-independent speaker verification,''
\newblock in {\em Proc. Interspeech}, 2019.

\bibitem{wan2018generalized}
L. Wan, Q. Wang, A. Papir and I.~L. Moreno,
\newblock ``Generalized end-to-end loss for speaker verification,''
\newblock in {\em Proc. ICASSP}, 2018.

\bibitem{park2022review}
T.~J. Park et~al.,
\newblock ``A review of speaker diarization: Recent advances with deep learning,''
\newblock {\em Computer Speech \& Language}, vol. 72, 2022.

\bibitem{jung2023search}
J.-w. Jung, H.-S. Heo et~al.,
\newblock ``In search of strong embedding extractors for speaker diarisation,''
\newblock in {\em Proc. ICASSP}. IEEE, 2023.

\bibitem{YourTTS-Casanova2022}
E. Casanova et~al.,
\newblock ``{YourTTS}: Towards zero-shot multi-speaker {TTS} and zero-shot voice conversion for everyone,''
\newblock in {\em Proc. ICML}, 2022.

\bibitem{VoiceFilter-Wang2019}
Q. Wang et~al.,
\newblock ``{VoiceFilter}: Targeted voice separation by speaker-conditioned spectrogram masking,''
\newblock in {\em Proc. Interspeech}, 2019.

\bibitem{SpEx-Xu2020}
C. Xu, W. Rao et~al.,
\newblock ``{SpEx}: Multi-scale time domain speaker extraction network,''
\newblock {\em IEEE/ACM Trans. ASLP.}, vol. 28, 2020.

\bibitem{huang2021far}
T.-h. Huang, J.-h. Lin and H.-y. Lee,
\newblock ``How far are we from robust voice conversion: A survey,''
\newblock in {\em Proc. SLT}, 2021.

\bibitem{huang2023singing}
W.-C. Huang, L.~P. Violeta, S. Liu et~al.,
\newblock ``The singing voice conversion challenge 2023,''
\newblock {\em arXiv:2306.14422}, 2023.

\bibitem{Kaldi-Povey2011}
D. Povey, A. Ghoshal, G. Boulianne et~al.,
\newblock ``The {Kaldi} speech recognition toolkit,''
\newblock in {\em Proc. ASRU}, 2011.

\bibitem{chung2020defence}
J.~S. Chung, J. Huh, S. Mun et~al.,
\newblock ``In defence of metric learning for speaker recognition,''
\newblock in {\em Proc. Interspeech}, 2020.

\bibitem{wang2023wespeaker}
H. Wang et~al.,
\newblock ``Wespeaker: A research and production oriented speaker embedding learning toolkit,''
\newblock in {\em Proc. ICASSP}, 2023.

\bibitem{tong2021asv}
F. Tong, M. Zhao, J. Zhou et~al.,
\newblock ``{ASV-Subtools}: {Open} source toolkit for automatic speaker verification,''
\newblock in {\em Proc. ICASSP}, 2021.

\bibitem{zhang2022paddlespeech}
H. Zhang, T. Yuan, J. Chen et~al.,
\newblock ``{P}addle{S}peech: An easy-to-use all-in-one speech toolkit,''
\newblock in {\em Proc. NAACL}, 2022.

\bibitem{ravanelli2021speechbrain}
M. Ravanelli, T. Parcollet, P. Plantinga et~al.,
\newblock ``{SpeechBrain}: A general-purpose speech toolkit,''
\newblock {\em arXiv:2106.04624}, 2021.

\bibitem{kuchaiev2019nemo}
O. Kuchaiev, J. Li, H. Nguyen et~al.,
\newblock ``{NeMo}: a toolkit for building ai applications using neural modules,''
\newblock {\em arXiv:1909.09577}, 2019.

\bibitem{zheng20233d}
S. Zheng, L. Cheng et~al.,
\newblock ``{3D-Speaker}: A large-scale multi-device, multi-distance, and multi-dialect corpus for speech representation disentanglement,''
\newblock {\em arXiv:2306.15354}, 2023.

\bibitem{hayashi2021espnet2}
T. Hayashi, R. Yamamoto, T. Yoshimura et~al.,
\newblock ``{ESP}net2-{TTS}: Extending the edge of {TTS} research,''
\newblock {\em arXiv:2110.07840}, 2021.

\bibitem{yang21c_interspeech}
S. wen Yang et~al.,
\newblock ``{SUPERB: Speech Processing Universal PERformance Benchmark},''
\newblock in {\em Proc. Interspeech}, 2021.

\bibitem{chen2022wavlm}
S. Chen, C. Wang et~al.,
\newblock ``{WavLM}: Large-scale self-supervised pre-training for full stack speech processing,''
\newblock {\em IEEE Journal of Selected Topics in Signal Processing}, vol. 16, no. 6, 2022.

\bibitem{mun2023frequency}
S.~H. Mun, J.-w. Jung et~al.,
\newblock ``Frequency and multi-scale selective kernel attention for speaker verification,''
\newblock in {\em Proc. SLT}, 2022.

\bibitem{jung2022sasv}
J.-w. Jung et~al.,
\newblock ``{SASV} 2022: The first spoofing-aware speaker verification challenge,''
\newblock in {\em Proc. Interspeech}, 2022.

\bibitem{desplanques2020ecapa}
B. Desplanques, J. Thienpondt et~al.,
\newblock ``{ECAPA-TDNN}: Emphasized channel attention, propagation and aggregation in {TDNN} based speaker verification,''
\newblock in {\em Proc. Interspeech}, 2020.

\bibitem{Audio-Ko2015}
T. Ko, V. Peddinti, D. Povey and S. Khudanpur,
\newblock ``Audio augmentation for speech recognition,''
\newblock in {\em Proc. Interspeech}, 2015.

\bibitem{watanabe2018espnet}
S. Watanabe, T. Hori, S. Karita et~al.,
\newblock ``{ESPnet}: End-to-end speech processing toolkit,''
\newblock in {\em Proc. Interspeech}, 2018.

\bibitem{yamamoto2019speaker}
H. Yamamoto et~al.,
\newblock ``Speaker augmentation and bandwidth extension for deep speaker embedding.,''
\newblock in {\em Proc. Interspeech}, 2019.

\bibitem{chen2023build}
Z. Chen et~al.,
\newblock ``Build a {SRE} challenge system: Lessons from {VoxSRC} 2022 and {CNSRC} 2022,''
\newblock in {\em Proc. Interspeech}, 2023.

\bibitem{zhang2022mfa}
Y. Zhang, Z. Lv, H. Wu et~al.,
\newblock ``{MFA-Conformer}: Multi-scale feature aggregation conformer for automatic speaker verification,''
\newblock in {\em Proc. Interspeech}, 2022.

\bibitem{jung2022pushing}
J.-w. Jung, Y. Kim, H.-S. Heo et~al.,
\newblock ``Pushing the limits of raw waveform speaker recognition,''
\newblock in {\em Proc. Interspeech}, 2022.

\bibitem{okabe2018attentive}
K. Okabe, T. Koshinaka and K. Shinoda,
\newblock ``Attentive statistics pooling for deep speaker embedding,''
\newblock in {\em Proc. Interspeech}, 2018.

\bibitem{ioffe2015batch}
S. Ioffe and C. Szegedy,
\newblock ``Batch normalization: Accelerating deep network training by reducing internal covariate shift,''
\newblock in {\em Proc. ICML}, 2015.

\bibitem{ba2016layer}
J.~L. Ba, J.~R. Kiros and G.~E. Hinton,
\newblock ``Layer normalization,''
\newblock in {\em NeuralIPS - Deep Learning Symposium}, 2016.

\bibitem{deng2019arcface}
J. Deng et~al.,
\newblock ``{ArcFace}: Additive angular margin loss for deep face recognition,''
\newblock in {\em Proc. CVPR}, 2019.

\bibitem{zhao2021speakin}
M. Zhao et~al.,
\newblock ``The {SpeakIn} system for {VoxCeleb} speaker recognition challange 2021,''
\newblock {\em arXiv:2109.01989}, 2021.

\bibitem{matejka2017analysis}
P. Matejka et~al.,
\newblock ``Analysis of score normalization in multilingual speaker recognition.,''
\newblock in {\em Proc. Interspeech}, 2017.

\bibitem{torgashov2023id}
N. Torgashov et~al.,
\newblock ``The {ID} {R\&D} {VoxCeleb} speaker recognition challenge 2023 system description,''
\newblock {\em arXiv:2308.08294}, 2023.

\bibitem{nagrani2017voxceleb}
A. Nagrani, J.~S. Chung and A. Zisserman,
\newblock ``{VoxCeleb}: A large-scale speaker identification dataset,''
\newblock in {\em Proc. Interspeech}, 2017.

\bibitem{chung2018voxceleb2}
J.~S. Chung, A. Nagrani and A. Zisserman,
\newblock ``{VoxCeleb2}: Deep speaker recognition,''
\newblock in {\em Proc. Interspeech}, 2018.

\bibitem{lin2024voxblink}
Y. Lin, X. Qin, G. Zhao et~al.,
\newblock ``{VoxBlink}: A large scale speaker verification dataset on camera,''
\newblock in {\em Proc. ICASSP}, 2024.

\bibitem{wang2020asvspoof}
X. Wang, J. Yamagishi, M. Todisco et~al.,
\newblock ``{ASVspoof} 2019: A large-scale public database of synthesized, converted and replayed speech,''
\newblock {\em Computer Speech \& Language}, vol. 64, 2020.

\bibitem{jung2022aasist}
J.-w. Jung et~al.,
\newblock ``Aasist: Audio anti-spoofing using integrated spectro-temporal graph attention networks,''
\newblock in {\em Proc. ICASSP}, 2022.

\bibitem{kim2021conditional}
J. Kim et~al.,
\newblock ``Conditional variational autoencoder with adversarial learning for end-to-end text-to-speech,''
\newblock in {\em Proc. ICML}, 2021.

\bibitem{veaux201x_cstr_vctk}
C. Veaux, J. Yamagishi and K. MacDonald,
\newblock ``{CSTR VCTK Corpus: English Multi-speaker Corpus for CSTR Voice Cloning Toolkit},'' CSTR, University of Edinburgh, 2019.

\bibitem{radford2023robust}
A. Radford, J.~W. Kim, T. Xu et~al.,
\newblock ``Robust speech recognition via large-scale weak supervision,''
\newblock in {\em Proc. ICML}. PMLR, 2023.

\bibitem{saeki22c_interspeech}
T. Saeki, D. Xin et~al.,
\newblock ``{UTMOS: UTokyo-SaruLab System for VoiceMOS Challenge 2022},''
\newblock in {\em Proc. Interspeech}, 2022.

\bibitem{Deep-Hershey2016}
J.~R. Hershey et~al.,
\newblock ``Deep clustering: Discriminative embeddings for segmentation and separation,''
\newblock in {\em Proc. ICASSP}, 2016.

\bibitem{Improving-Delcroix2020}
M. Delcroix, T. Ochiai, K. Zmolikova et~al.,
\newblock ``Improving speaker discrimination of target speech extraction with time-domain {SpeakerBeam},''
\newblock in {\em Proc. ICASSP}, 2020.

\bibitem{zhao2022probing}
Z. Zhao et~al.,
\newblock ``Probing deep speaker embeddings for speaker-related tasks,''
\newblock {\em arXiv preprint arXiv:2212.07068}, 2022.

\bibitem{liu2023quantitative}
X. Liu, X. Li and J. Serr{\`a},
\newblock ``Quantitative evidence on overlooked aspects of enrollment speaker embeddings for target speaker separation,''
\newblock in {\em Proc. ICASSP}, 2023.

\bibitem{peng2023parameter}
J. Peng, T. Stafylakis, R. Gu et~al.,
\newblock ``Parameter-efficient transfer learning of pre-trained transformer models for speaker verification using adapters,''
\newblock in {\em Proc. ICASSP}, 2023.

\end{thebibliography}

\end{document}